\documentclass[a4paper]{article}
\usepackage{latexsym}
\begin{document}

\newtheorem{theorem}{Theorem}
\newtheorem{proposition}{Proposition}
\newtheorem{remark}{Remark}
\newtheorem{corollary}{Corollary}
\newtheorem{lemma}{Lemma}
\newtheorem{observation}{Observation}

\newcommand{\qed}{\hfill$\Box$\medskip}

\title{Counting Ones Without Broadword Operations} 
\author{Holger Petersen\\ 
Reinsburgstr. 75\\
70197 Stuttgart\\
Germany} 

\maketitle

\begin{abstract}
A lower time bound $\Omega(\min(\nu(x), n-\nu(x))$ for counting the number of
ones in a binary input word $x$ of length $n$ corresponding to the word length
of a processor architecture is presented, where $\nu(x)$ is
the number of ones. The operations available are increment, decrement, bit-wise
logical operations, and assignment. The only constant available is zero. An
almost matching upper bound is also obtained. 
\end{abstract}

\section{Introduction}
Counting ones among the $n$ bits of a machine word (also known as sideways addition 
\cite{WWG57,Knuth09} or bit count \cite{KR78}) is an operation that 
received considerable attention and has been implemented in several processor 
architectures including CDC 6600 in the 1960's, Cray-1 supercomputers in the 1970's (as population count),
and the ARM architecture Advanced SIMD (NEON) extensions (as vector count set bits VCNT)
in the new millennium.

A straight-forward method for counting ones (not making use of powerful operations 
like multiplication or division as in \cite{HAKMEM,WWG57} or broadword operations 
addition and shift \cite{Knuth09})
is to inspect every bit, resulting in complexity $O(n)$. An interesting improvement
was presented by Wegner \cite{Wegner60} based on the fact that for a non-zero
$x$ forming the expression $x\mbox{ AND } (x-1)$ deletes the right-most one. 
Repeating this process results in an algorithm of complexity $O(\nu(x))$, where $\nu(x)$ is the 
number of ones in the input $x$. This technique is also mentioned by 
(and sometimes attributed to)
Kernighan and Ritchie in their classical text-book \cite[Exercise~2-9]{KR78}. 
A similar approach using the twos complement of $x$ as a bit mask of a test instruction
is presented in \cite[p.~93]{DEC}\footnote{The powerful PDP-10 instruction TDZE tests its
operand with a bit mask and sets the selected bits to zero. It then skips the next instruction
if all selected bits were originally zero. Notice that the lower bits of $x$ and $-x$ are
identical up to and including the right-most one, while the remaining bits are complemented.
Instruction TDZE thus deletes this one when applied to $x$ and $-x$. 
The number of times TDZE can be applied before the operand becomes zero is $\nu(x)$.}.

In \cite{Petersen11} we pointed out, that we were not able to prove a lower bound 
$\Omega(\nu(x))$ matching the upper bound from \cite{Wegner60}
for algorithms based on the operations increment, decrement, and logical operations 
employed in Wegner's method. As will be shown in this note, such a bound
is not feasible, since for densely populated input words the complexity can be reduced. Densely
populated inputs are mentioned at the end of  \cite{Wegner60} and it is pointed out that
zeroes can be counted in the ones complement of the original input. 
By the remark after (63) in \cite{Knuth09} the improvement can be achieved 
without complementation.

\section{Preliminaries}\label{Preliminaries}
By  $\nu(x)$ we denote  the number of ones in $x$. 
We assume a computational model with unsigned integer variables. The input is stored in $x$
and all other variables are initially zero. The operations available are increment, decrement, logical operations 
AND, OR, and assignment. We assume that incrementing the
value consisting entirely of binary ones results in zero and decrementing 
zero results in this value. The only constant available in assignments and comparisons
is zero. 

\section{Results}

\begin{lemma}\label{Prefix}
Let $x = e(01)^m d^{n-2m-1}$ with $e, d\in \{ 0, 1\}$ and $0 \le m < n/2$ be an input of a program
as described in Section~\ref{Preliminaries}. We define $k_0 = n$ and 
$k_i = 2(m-i) + 1$ for $i  \ge 1$. After $i$ increment and decrement operations,
every variable satisfies the property that the $k_i$ most significant bits of its value are in the
set $\{ 0^{k_i}, 1^{k_i},  x' \}$ where $x'$ consists of the $k_i $ most significant bits of 
input $x$.
\end{lemma}
{\bf Proof.} 
The claim trivially holds for the initial state of the program since all variables except for $x$
are zero. As long as no increment and decrement operations are carried out, values satisfying
the claim can be transferred or a variable can be set to zero, which preserves the claimed property. 
If the operands of operation AND are in the set
$\{ 1^n, x \}$ the result will be in the same set and it will be  $0^n$ otherwise. 
Symmetrically OR will result in  $x$ or $0^n$ if the 
operands are in the set $\{ 0^n, x \}$ and $1^n$ otherwise. This shows the claim for $i = 0$.
Suppose that $i > 0$ and the claim holds for $i-1$. Again, assignments can only transfer values satisfying
the claim (observe that the restriction for $i$ is weaker than for $i-1$) or set all bits of a variable
to zeroes, which leads to a string in the set. Logical operation AND 
will result in prefix $0^{k_i}$ if $0^{k_i}$ is among the prefixes of the operands and 
will leave the other operand unchanged if $1^{k_i}$ is among the prefixes.
Operation OR results in $1^{k_i}$ if $1^{k_i}$ is among the prefixes of the operands
and will leave the other operand unchanged if $0^{k_i}$ is among the prefixes.
Combining $e(01)^{(k_i-1)/2}$ with itself results in the same string for both logical operations. 
The prefixes of length $k_i$ thus form three-element monoids with these operations.
Now consider the increment operation. If there is no carry into the leading $k_{i-1}$ bits, then  
the leading $k_{i}$ bits are in the set by assumption. 
If there is a carry, then either all leading bits are 1 and 
they are switched to 0, or at least bit $k_i-1$ is equal to 0 and thus the carry cannot propagate.
Similarly, for decrement either $0^{k_i}$ changes to $1^{k_i}$ or the borrow cannot 
propagate into the leading bits because bit $k_i$ is equal to 1. This shows the claim for the
$i$-th increment or decrement operation. After this operation an argument as above
shows that the claim is preserved by the other operations. \qed

\begin{theorem}\label{lowerWegner}
Counting ones in a word $x$ of length $n > 1$ requires $\min(\nu(x), n-\nu(x))$ steps 
for  $\nu(x) \neq n/2$ in the worst case under 
unit cost measure using the operations increment, decrement, AND, OR,
and assignment where $\nu(x)$ is the number of ones in the input $x$.
\end{theorem}
{\bf Proof.} We consider inputs $x$ of the form $1(01)^m 1^{n-2m-1}$ or $0(01)^m 0^{n-2m-1}$ where
$0 \le m < n/2$. Notice that inputs of the first form cover the range $\nu(x) > n/2$ and 
the second form covers $\nu(x) < n/2$. By Lemma~\ref{Prefix} after $i$ operations 
the variables have prefixes in the set $\{ 0^{k_i}, 1^{k_i}, d(01)^{(k_i-1)/2} \}$ with $d \in \{ 0, 1\}$.

We assume that an algorithm can compute $\nu(x) \neq n/2$ with less than $\min(\nu(x), n-\nu(x))$
increment and decrement operations. We modify the input $x$ by replacing the most significant bit
$d$ with $d' = 1-d$ and claim that the algorithm performs the same sequence of 
operations on this modified input $x'$ as on $x$ and 
that therefore all corresponding variables have the same values except for possibly the most 
significant bit. The flow of control can only differ if a comparison has a different result. If a variable
with a prefix $000$ is compared with $001$ or $101$ it will be smaller independently of the lower
bits. Similarly $111$ is larger than $001$ or $101$ and the least significant bits are
irrelevant. Finally a comparison of variables with the 
same prefix depends on the lower order bits, which will be identical by assumption.

If a variable having a value with prefix $000$ or $111$ is output, we obtain a contradiction since 
$\nu(x') \neq \nu(x)$ and the output is the same on both inputs. 
If a variable with the most significant bit $d'$ is output, its value 
is at least $\nu(x)+2$ if $d = 0$ or at most $\nu(x)-2$ if $d = 1$ because $n >1$.
Again this leads to a contradiction, since $\nu(x') \in \{ \nu(x)-1, \nu(x)+1 \}$.\qed
 
Now we investigate upper bounds on computing $\nu(x)$. 
For a single bit the input represents the count of ones and no operations are required. For two
bits a single decrement suffices to count ones:

\begin{observation}
Counting ones in a word $x$ of length two can be done with one decrement operation
if the input $x$ is not zero and without increment or decrement operation otherwise. 
This bound cannot be improved.
\end{observation}
{\bf Proof.} The following algorithm in C shows the upper bound:
\begin{verbatim}
int countones(int x)
{
    int y;
    
    if (x == 0) return 0;
    else
    {
        y = x-1;
        if (y == 0) return x;
        else return y;
    }
}
\end{verbatim}
Suppose for an input $x\in\{ 01, 10, 11 \}$ the computation of $\nu(x)$ does not require
an increment or decrement operation. By Lemma~\ref{Prefix} the output is in $\{ 00, 11, x \}$
and therefore the algorithm cannot work correctly for $10$ and $11$. The input $01$ could be 
output as the result, but then the computation for input $11$ will result in the incorrect 
output $11$. \qed

For input lengths larger than two it seems to be necessary to build up a count 
of ones or count down from $n$ while processing the input. We first show that
the value $n$ can be generated quite efficiently and then use this construction for
counting ones in densely populated inputs.

\begin{lemma}\label{constantWegner}
A constant $n$ can be generated in $O(\log n)$ steps starting from zero
using increment and the logical operation OR.
\end{lemma}
{\bf Proof.} Let $n = 2^{k_1} + 2^{k_2} + \cdots + 2^{k_m}$ with ${k_1} < {k_2} <  \cdots < {k_m}$.
Notice that $n = 2^{k_1} \mbox{ OR }  2^{k_2} \mbox{ OR } \cdots \mbox{ OR } 2^{k_m}$ as well.
Let $t_{1} = 0$ and
generate  $t_{i+1} = 2^{i+1}-1$ from $t_i = 2^{i}-1$ for $i>1$
by forming $t_{i+1} = t_{i} \mbox{ OR } t_{i}+1$.
By selecting the suitable numbers we form $n = (t_{k_1}+1) \mbox{ OR } ( t_{k_2}+1) \mbox{ OR } \cdots \mbox{ OR } 
(t_{k_m}+1)$.\qed

\begin{theorem}\label{upperWegner}
Counting ones in words of length $n$ can be done in $O(\min(\nu(x), n-\nu(x)+\log n))$ steps under 
unit cost measure using the operations increment, decrement, logical operations 
and assignment of zero.
\end{theorem}
{\bf Proof.} We will describe below an algorithm for densely populated inputs that has complexity 
$O(n-\nu(x)+\log n)$. This solution will be run in parallel on a copy of the input
with Wegner's method interleaving steps of the two approaches. As soon as one of the algorithms stops, 
its output is the result of the combined method. Clearly this gives the claimed bound.

By Lemma~\ref{constantWegner} we may initialize a counter $b$ with the length $n$
in $O(\log n)$ steps. Then we start
forming $x \mbox{ OR }  (x+1)$ as long as $x$ has not reached the maximum value. The latter
can be detected by comparing $x+1$ with zero. For every iteration we decrement $b$. Since 
the number of iterations is equal to
the number of zeroes, the resulting value of $b$ is the number of ones. \qed

\section{Discussion}

The following table summarizes lower and upper time bounds for counting ones 
with different sets of operations.

\begin{center}
\begin{tabular}{|l|c|c|}\hline
set of operations               & lower bound  & upper bound  \\\hline\hline
increment, decrement, &  $\Omega(\min(\nu(x),$ & $O(\min(\nu(x),$  \\
AND, OR  &  $\  n-\nu(x))$ & $\  n-\nu(x)+\log n))$  \\
only constant 0  & (Theorem~\ref{lowerWegner}) &  (Theorem~\ref{upperWegner}) \\\hline
addition, AND, OR    & $\Omega(\log n/\log\log n)$ & $O(\log^2n)$ \\
(PAL)   &  \cite[Exercise~127]{Knuth09} &  \cite{Petersen11}    \\\hline
addition, shift, AND, OR &  $\Omega(\log n/\log\log n)$  & $O(\log n)$  \\
(broadword steps) & \cite[Exercise~127]{Knuth09}  & (folklore, see \cite{Chess})   \\\hline
addition, shift, AND, OR, &  & $O(\log^*n)$   \\
multiplication  &  &  \cite{Petersen11}  \\\hline
addition, shift, AND, OR, & & $O(\log\log n)$   \\
division  &  &  \cite[Item~169]{HAKMEM}    \\\hline
\end{tabular}
\end{center}

The lower time bound $\Omega(\log n/\log\log n)$ on parity (and thus counting ones) 
in the model of broadword steps follows from a result in circuit complexity.



\end{document}